\documentclass[12pt]{emulateapj}
\usepackage{color}
\usepackage{threeparttable}

\newcommand\ha{H$\alpha$~}

\shorttitle{Evidence for a gas rich major merger}
\shortauthors{Tadaki et al.}

\begin{document}


\title{Evidence for a gas-rich major merger in a proto-cluster at $z$=2.5}


\author{Ken-ichi Tadaki\altaffilmark{1}, Tadayuki Kodama\altaffilmark{1,2}, Yoichi Tamura\altaffilmark{3}, Masao Hayashi\altaffilmark{1,4}, Yusei Koyama\altaffilmark{1,5}, Rhythm Shimakawa\altaffilmark{2}\\
Ichi Tanaka\altaffilmark{6}, Kotaro Kohno\altaffilmark{3,7}, Bunyo Hatsukade\altaffilmark{1}, and Kenta Suzuki\altaffilmark{3}
}


\affil{\altaffilmark{1}National Astronomical Observatory of Japan, Mitaka, Tokyo 181-8588, Japan; tadaki.ken@nao.ac.jp}
\affil{\altaffilmark{2}Department of Astronomical Science, The Graduate University for Advanced Studies, Mitaka, Tokyo 181-8588, Japan}
\affil{\altaffilmark{3}Institute of Astronomy, The University of Tokyo, Mitaka, Tokyo 181-0015, Japan}
\affil{\altaffilmark{4}Institute for Cosmic Ray Research, The University of Tokyo, 5-1-5 Kashiwanoha, Kashiwa, Chiba 277-8582, Japan}
\affil{\altaffilmark{5}Institute of Space Astronomical Science, Japan Aerospace Exploration Agency, Sagamihara, Kanagawa 252-5210, Japan}
\affil{\altaffilmark{6}Subaru Telescope, National Astronomical Observatory of Japan, 650 North A'ohoku Place, Hilo, HI 96720, USA}
\affil{\altaffilmark{7}Research Center for the Early Universe, The University of Tokyo, 7-3-1 Hongo, Bunkyo, Tokyo 113-0033, Japan}


\begin{abstract}
Gas-rich major mergers in high-redshift proto-clusters are important events, perhaps leading to the creation of the slowly rotating remnants seen in the cores of clusters in the present day.
Here, we present a deep Jansky Very Large Array observation of CO $J$ = 1--0 emission line in a proto-cluster at $z = 2.5$, USS1558-003. The target field is an extremely dense region, where 20 H$\alpha$ emitters (HAEs) are clustering. 
We have successfully detected the CO emission line from three HAEs and discovered a close pair of red and blue CO-emitting HAEs. 
Given their close proximity ($\sim$30 kpc), small velocity offset ($\sim$300 km s$^{-1}$), and similar stellar masses, they could be in the early phase of a gas-rich major merger. 
For the red HAE, we derive a total infrared luminosity of $L_\mathrm{IR}=5.1\times10^{12}~L_\odot$ using MIPS 24 $\mu$m and radio continuum images. 
The $L_\mathrm{IR}/L'_\mathrm{CO}$ ratio is significantly enhanced compared to local spirals and high-redshift disks with a similar CO luminosity, which is indicative of a starburst mode. 
We find the gas depletion timescale is shorter than that of normal star-forming galaxies regardless of adopted CO-H$_2$ conversion factors.
The identification of such a rare event suggests that gas-rich major mergers frequently take place in proto-clusters at $z > 2$ and may involve the formation processes of slow rotators seen in local massive clusters.
\end{abstract}





\section{Introduction}
\label{sec;intro}

Properties of galaxies seen in the local universe strongly depend on their surrounding environments, as is well known from the morphology--density relation. 
Early-type galaxies are frequently observed in high-density regions such as clusters and late-type galaxies are common in low-density regions \citep{1997ApJ...490..577D}.
The ATLAS$^\mathrm{3D}$ survey further demonstrates that early-type galaxies are divided into two kinds of populations based on the kinematics: fast rotators and slow rotators \citep{2011MNRAS.414..888E}.
This categorization makes the trend in the morphology--density relation more prominent.
Whereas fast rotators, which form the majority of early-type galaxies, appear in a wide range of environments, slow rotators reside exclusively in dense cores of mature clusters \citep{2011MNRAS.416.1680C}.

Such spatial segregation and the difference in their kinematics could be related to formation processes and subsequent quenching mechanisms of star formation, which probably take place at high redshift.
In theoretical models, gas-rich major mergers leading to a spin-down of remnants successfully produce simulated slow rotators and are considered to be one of the possible formation processes of slow rotators \citep{2013arXiv1311.0284N}. 
Given high number densities of star-forming galaxies in proto-clusters at $z>2$ \citep[e.g.][]{2011PASJ...63S.415T}, 
we naturally expect a high frequency of major-merger events.
What is important here is whether they are interactions between gas-dominated systems.
While stellar components within galaxies are collisionless, systems comprising gas are dissipational.
Therefore, gas-rich major mergers trigger an intense, dusty star formation due to shocks and an inflow of gas that has lost its angular momentum, as well as establishment of the outer profile through violent relaxation \citep{1996ApJ...464..641M}. 

CO observations are critical for measuring the molecular gas mass within galaxies, $M_\mathrm{gas}$, and investigating the star-formation mode characterized by the star-formation efficiency, SFE=SFR/$M_\mathrm{gas}$, and the gas depletion timescale, $\tau_\mathrm{depl}=M_\mathrm{gas}$/SFR.
CO studies at high-redshift have rapidly developed over the past years not only for very bright galaxies in the dust emission such as submillimeter galaxies (SMGs; \citealt{1998ApJ...506L...7F, 2003ApJ...597L.113N, 2005MNRAS.359.1165G, 2008ApJ...680..246T, 2010ApJ...724..233E}) but for optical/near-infrared selected galaxies such as $BzK$ galaxies \citep[e.g.][]{2010ApJ...713..686D, 2013ApJ...768...74T}.
However, most of them observe high-excitation CO lines mainly using the Plateau de Bure Interferometer.
This significantly affects the estimates of molecular gas mass because high-$J$ lines trace dense gas regions rather than total gas reservoirs probed by CO $J=1-0$ emission.

In a cluster field at $z=0.4$, \cite{2011ApJ...730L..19G} detect the CO $J=1-0$ line from five dusty star-forming galaxies and find the environmental dependence of SFE is not seen. 
\cite{2012MNRAS.426..258A} also find that the SFEs of two galaxies in a proto-cluster at $z=1.5$ are comparable to that of field galaxies at similar redshift.
On the other hand, \cite{2013ApJ...772..137I} discover four CO luminous galaxies across an $\sim$100 kpc region at $z=2.4$ and find that two of them have a high SFE.
No conclusive result could be obtained due to a small sample size.
In this paper, we report results from a deep CO $J=1-0$ observation of a proto-cluster at $z=2.5$ to search for gas-rich galaxies and see if there is any environmental effect of the star-formation mode in the formation phase of the progenitors of cluster early-type galaxies seen today.
We assume the Salpeter initial mass function \citep{1955ApJ...121..161S} and cosmological parameters of H$_0$ = 70 km s$^{-1}$ Mpc$^{-1}$, $\Omega _\mathrm{M}$ = 0.3, and $\Omega _\Lambda$ = 0.7. 

\begin{figure}
\begin{center}
\includegraphics[width=0.9\linewidth]{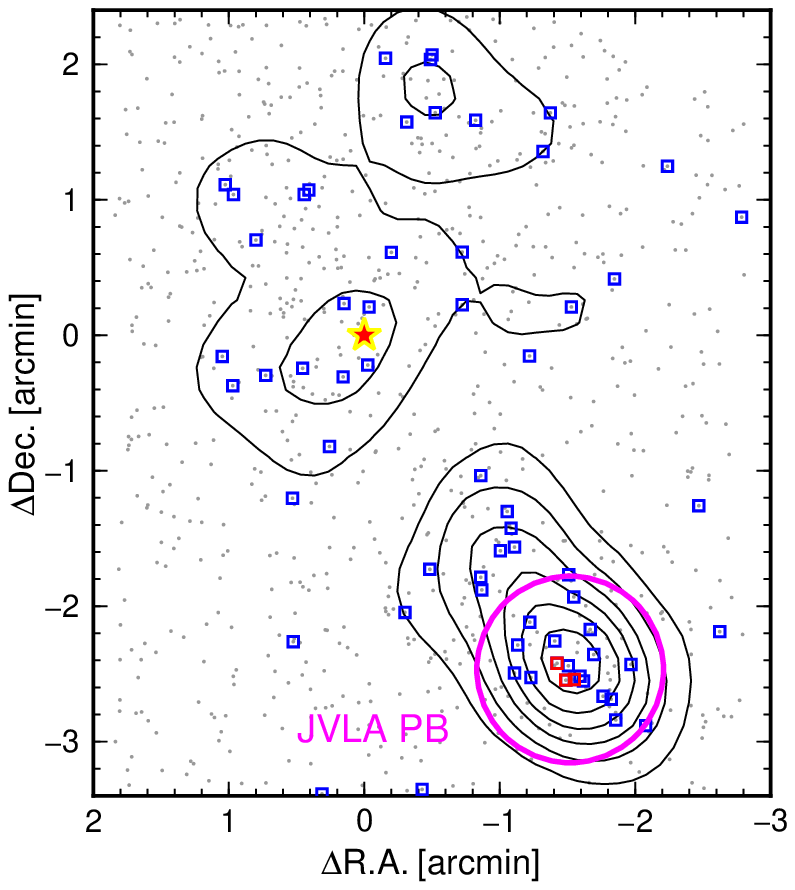}
\caption{A 2-D map of the USS1558 proto-cluster at $z = 2.5$. 
Red and blue squares represent CO-detected and non-detected \ha emitters (HAEs), respectively. 
Coordinates are relative to the radio galaxy (a red star). 
The primary beam is shown by a magenta circle. Contours denote the local number densities ($\Sigma_\mathrm{5th}$) of all HAEs, in a step of 2 Mpc$^{-2}$. \label{fig;distribution}}
\end{center}
\end{figure}

\section{Observation}

\subsection{USS1558-003 proto-cluster at $z=2.5$}

Our target is a proto-cluster at $z=2.5$, USS1558-003, where an over-density of massive, red-sequence galaxies has been discovered around a radio galaxy \citep{2007MNRAS.377.1717K}.
This region has also been observed as part of a systematic \ha narrow-band imaging campaign with MOIRCS on Subaru Telescope called ``$MAHALO-Subaru$'' project (MApping HAlpha and Lines of Oxygen with Subaru; \citealt{2013IAUS..295...74K}), and it is found to host numerous \ha emitters (HAEs; \citealt{2012ApJ...757...15H}).
Membership of about half of the 68 HAEs has been spectroscopically confirmed with a success rate of 70\% \citep{2014MNRAS.441L...1S}.
An extremely dense clump lies about three arc-minutes away from a radio galaxy in the southwest (Figure \ref{fig;distribution}).
Because the dynamical mass of this clump is estimated to be  $\sim10^{14}~M_\odot$ from the \ha spectroscopy,
it is expected to evolve into a single massive system with $>10^{15}~M_\odot$ similar to the Coma cluster \citep{2014MNRAS.441L...1S}.
Therefore, this proto-cluster is very likely the site where slow rotators, which will eventually dominate a rich cluster by the present-day, are just in their formation phase.

\subsection{JVLA observations}

We have conducted CO $J=1-0$ emission line observations with the Jansky Very Large Array (JVLA) during February-April 2013.
The target field includes 20 HAEs at $z=2.5$ (Figure \ref{fig;distribution}), of which 12 have spectroscopic redshifts based on their \ha line.
The observations were made in the compact D array configuration to securely detect emission lines from entire galaxies.
Because the narrow-band survey samples star-forming galaxies at $z=2.53\pm0.02$, 
the CO $J=1-0$ emission line ($\nu_\mathrm{rest}=115.271$ GHz) can be observed with the Ka-band receiver ($\nu_\mathrm{obs}=33$ GHz), providing the primary beam size of 82\arcsec.
The WIDAR correlator was set up to cover 32.078--34.082 GHz, corresponding to
the CO line at $z$=2.382--2.593.
We observed the standard calibrators 3C286 (1.9 Jy) for bandpass and flux calibration.
Phase calibration is performed with observations of J1557-0001 (0.7 Jy).
The data are processed through the VLA CASA Calibration Pipeline \citep{2007ASPC..376..127M}.
Channels at the edges of each spectral window are flagged, which brings 14 spectral gaps of 14 MHz.
The total integration time is 14 hours on source.
Two kinds of CO maps are reconstructed with the CASA task CLEAN, using ``briggs'' weighting with a robustness parameter of -0.5 and natural weighting with 5.0, which provide a synthesized beam of 2.2$\times$1.9 arcsec$^2$/3.0$\times$2.5 arcsec$^2$ and a rms level of 80--90 $\mu$Jy beam$^{-1}$/40--50 $\mu$Jy beam$^{-1}$ per 2 MHz channel (18 km s$^{-1}$) before the primary beam correction, respectively.

\begin{figure*}[t]
\includegraphics[width=1.0\linewidth]{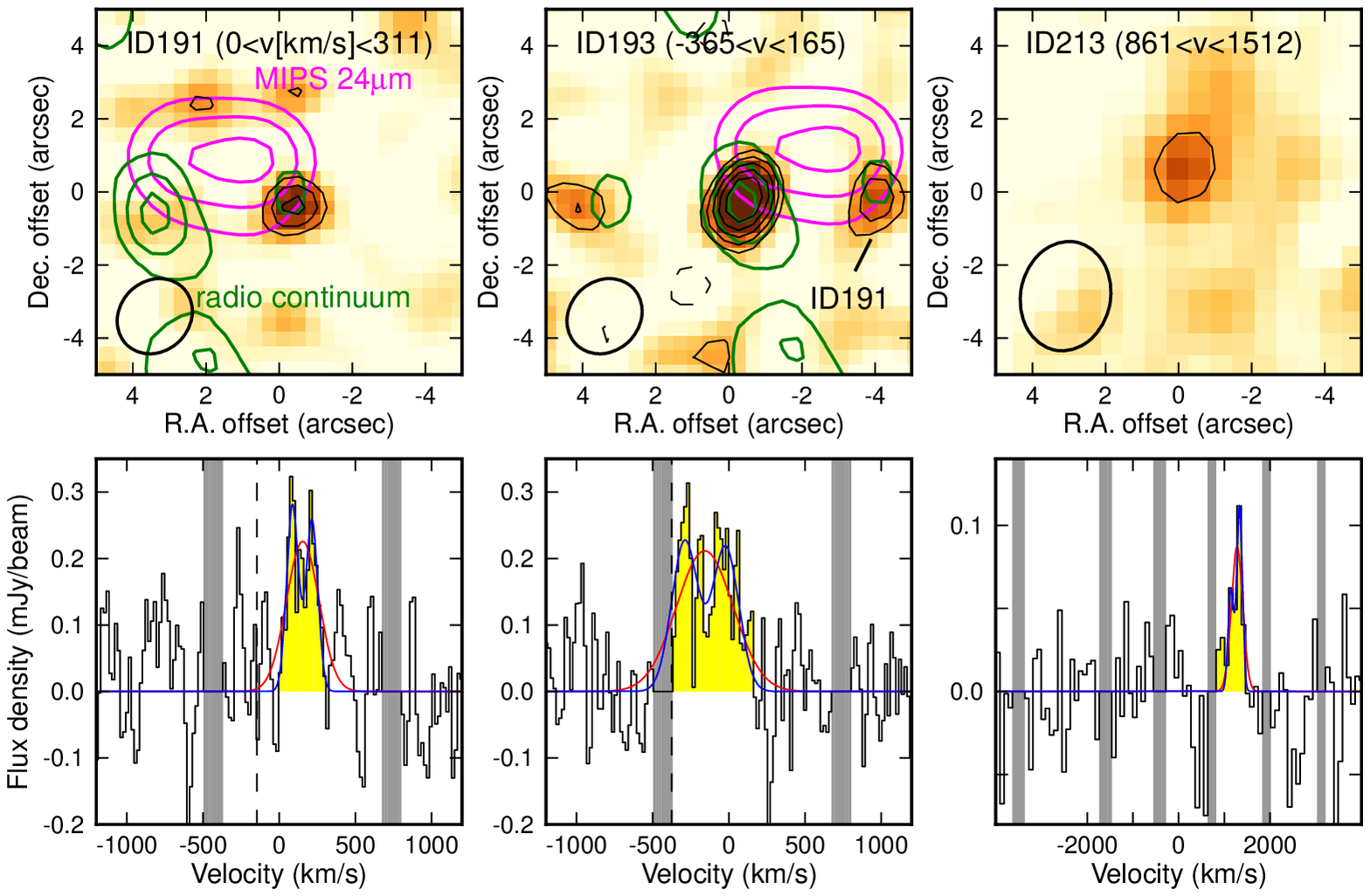}
\caption{(Top) Velocity-averaged maps of the CO emission for three HAEs. 
The visibility data are averaged over the spectral bins where positive signals are detected (yellow-shaded regions in the bottom panel).
The images are obtained using briggs weighting for ID 191/ID 193 and natural weighting for ID 213. 
The beamsize is at bottom left .
Magenta and green contours present MIPS 24 $\mu$m and radio continuum ($\nu_\mathrm{rest}=116$ GHz) map, respectively.
(Bottom) The CO spectra extracted from the peak position in the top images.
The horizontal axis shows the relativistic velocity in the rest frame at $z=2.515$.
The velocity resolution is 18 km s$^{-1}$, but the spectrum of ID 213 is binned over the velocity range of 92 km s$^{-1}$ to detect a faint emission line.
The red and blue lines show the best-fitting profile with single and double gaussian model, respectively.
Dashed lines indicate the redshift expected from the \ha spectroscopy.
The channels flagged at the edges of each spectral window are shown by gray-shaded regions.
\label{fig;co_map}}
\end{figure*}

\subsection{Spitzer MIPS 24 $\mu$m data}

We use a $Spitzer$/MIPS 24 $\mu$m image to identify dusty star formation and estimate total infrared (IR) luminosities of HAEs.
The data is retrieved from the Spitzer Heritage Archive.
Data reduction is performed in a standard manner (flat fielding, background subtraction and mosaic) using MOPEX software \citep{2005ASPC..347...81M}.
Source extraction and PSF-fitted photometry are performed using APEX module in MOPEX.
The limiting flux reaches to $5\sigma\sim150~\mu$Jy in the final combined image.

\section{Result}
\label{sec;result}

\subsection{Detections of CO J=1-0 emission line}
\label{sec;detection}

We search for the CO $J=1-0$ emission line within a 0.5 arcsec radius from the position of 20 HAEs by using the 10 MHz ($\sim$92 km s$^{-1}$) binned data cube.
To avoid spurious detections, we identify a $>3\sigma$ peak associated with two additional bins of $>1.5\sigma$ significance.
The probability that such signals are detected, by chance, in the frequency range (32.471--32.841 GHz, 37 bins) of the narrow-band redshift is estimated to be 0.5\% in Gaussian noise.
Actually, negative peaks are not seen except for one object located at the edge of the primary beam.
The significance of the detection is defined by the signal-noise ratio of peak flux in the 10MHz binned cube.
ID 191, ID 193, and ID213 are eventually detected in 6.9, 6.3, and $3.8\sigma$, respectively.
The separation between ID 191 and ID 193 is about 4 arcsec corresponding to 32 kpc in the physical scale and the velocity offset is 300 km s$^{-1}$ (Figure \ref{fig;co_map}).
We also look into the relation between the detection rate of CO emission and the rest-frame optical color.
Our sample consists of four red HAEs with $J-K_s>1.38$ and 16 blue HAEs.
The two (ID 193 and ID 213) out of four red HAEs are actually detected in the CO emission line while only one (ID 191) out of 16 blue HAEs is detected.
Red and massive galaxies tend to be relatively bright in CO emission compared to blue and less massive ones.

\begin{table*}[t]
\begin{center}
\caption{Properties of the \ha emitters with CO(1-0) line detections}
\begin{threeparttable}
\begin{tabular}{lcccccccc}
\hline
ID$^\mathrm{a}$ & R.A. & Decl. & $z_\mathrm{CO}$ & $v_\mathrm{CO}$ FWHM $^\mathrm{b}$ & $S_\mathrm{CO}dv~^\mathrm{c}$ & $L'_\mathrm{CO}$ & $L_\mathrm{IR}$ & $M_*~^\mathrm{d}$ \\
& (J2000) & (J2000) & & (km s$^{-1}$) & (Jy km s$^{-1}$) & ($10^{10}$ K km s$^{-1}$pc$^2$) & ($10^{12}$ $L_\odot$) & ($10^{10}~M_\odot$)\\
\hline
bHAE-191& 16 01 11.20 & $-$00 31 19.0 & 2.5168 & 251 & 0.052 $\pm$ 0.008 & 1.5 & 2.5 & 5.1 \\
rHAE-193 & 16 01 11.45 & $-$00 31 19.3 & 2.5131 & 437 & 0.096 $\pm$ 0.015 & 2.8 & 5.1 & 4.8 \\
rHAE-213 & 16 01 11.71 & $-$00 31 11.9 & 2.5300 & 294 & 0.026 $\pm$ 0.007 & 0.8 & $<$1.7 & 4.6 \\
\hline
\end{tabular}
\begin{tablenotes}\footnotesize
\item[a] ``bHAE'' and ``rHAE'' indicate blue and red HAEs, respectively, separated at $J-K_s=1.38$.
\item[b] FWHM of the single gaussian component.
\item[c] Velocity-integrated flux derived from the two-component model. The errors are calculated on the basis of the signal-noise ratio defined by the 10 MHz binned cube.
\item[d] Stellar masses are estimated from $K_s$-band magnitudes using the relationship between mass-to-luminosity ratios in $K_s$-band and $J-K_s$ colors based on the population synthesis bulge-disk composite models \citep{1999MNRAS.302..152K}.
\end{tablenotes}
\end{threeparttable}
\end{center}
\end{table*}


Table 1 shows a summary of the properties of CO emission line.
We fit their spectra with a single and two-component gaussian model to measure the line width and the velocity-integrated flux, $S_\mathrm{CO}dv$ (Figure \ref{fig;co_map}).
The FWHM of CO $J=1-0$ line profiles is 250--450 km s$^{-1}$, which is significantly smaller than that of bright SMGs \citep{2011MNRAS.412.1913I,2012MNRAS.425.1320I,2013ApJ...772..137I} and consistent with that of submm faint galaxies like BzK \citep{2012MNRAS.426..258A}.
The non-binning spectra with 2 MHz resolution shows a double-peak profile.
This irregular feature is closely related to a spatial distribution of molecular gas within galaxies.
Local edge-on disks show a double-horn profile but a gap at zero-velocity is widespread in the velocity range and moreover rarely below 50\% of the peak flux \citep[e.g.][]{2008AJ....136.2563W}.
Such spectra are also seen in color-selected star-forming galaxies at $z\sim1.5$ \citep{2010ApJ...713..686D} and SMGs at $z=1-3.5$ \citep{2005MNRAS.359.1165G}.
\cite{2010ApJ...713..686D} demonstrate with numerical simulations that turbulent and clumpy disks can successfully reproduce the spectra similar to the observed ones but uniform disks cannot. 
Actually, morphologies of star-forming galaxies become clumpy at $z>2$ \citep[e.g.][]{2014ApJ...780...77T}. 
If most of the CO emission is dominated by a few giant clumps, the observed feature can be readily explained.
Otherwise, one double profile would reflect a merger between two galaxies with different velocities.
Even higher-resolution observations would make it challenging to discriminate between a merger and multiple kpc-scale clumps within a single galaxy.
We discuss further the physical process in terms of star-formation mode in Section \ref{sec;discussion}.

\subsection{Identification of dusty star formation}

Although \ha emission line is one of the best indicators of SFR, it can still miss much of star formation in the case of very dusty galaxies \citep{2010MNRAS.403.1611K}.
Infrared emission, which is a re-radiation of UV flux of massive stars by their surrounding dust, is very useful for estimating dust-obscured SFRs of galaxies.
Three MIPS 24 $\mu$m sources are identified within the primary beam, and 
the brightest one with $S_\mathrm{24 \mu m}=425~\mu$Jy is located between ID 191 and ID 193 (Figure \ref{fig;co_map}).
Since the PSF size of the MIPS images is too large ($\sim6$\arcsec), we cannot measure the 24 $\mu$m flux density of each galaxy.

We use a radio continuum image to deblend the IR emission, because it is a good indicator of extinction-free SFR reflecting the non-thermal emission from supernovae remnants and free-free emission from ionized gas \citep[e.g.][]{2011ApJ...737...67M,2012MNRAS.425.2203T}.
A 33 GHz (116 GHz in the rest frame) image is created with natural weighting by averaging the data over about 2 GHz excluding the frequency range of the CO emission line.
The radio data allow us to speculate the relative contribution of each galaxy to the IR emission.
We estimate a 116 GHz flux density of 4.8 $\pm~2.6~\mu$Jy (1.8$\sigma$) and 10.2 $\pm~2.6~\mu$Jy (3.9$\sigma$) for ID 191 and ID 193, respectively (Figure 2).
Although ID 191 is marginally detected, it is clear that the blended IR emission is dominated by ID 193.
\cite{2011ApJ...738..106W} demonstrate with $Hershel$/PACS data that the conversion from observed 24 $\mu$m photometries to total IR luminosities, $L_\mathrm{IR}$, is applicable with $\sim$0.3 dex scatter out to $z=3$.
Using the luminosity-independent conversion factor \citep{2008ApJ...682..985W}, we derive $L_\mathrm{IR}=2.5\times10^{12}~L_\odot$ and $L_\mathrm{IR}=5.1\times10^{12}~L_\odot$ for ID 191 and ID 193, respectively.
Assuming that the bulk of the total IR luminosity is powered by star formation, we find with the standard calibration of \cite{1998ARA&A..36..189K} that ID 193 has a high SFR of $886$ M$_\odot$yr$^{-1}$.
Since the H$\alpha$-based SFR is $\sim$40 M$_\odot$yr$^{-1}$, this object is likely to be strongly attenuated by a large amount of dust (A$_\mathrm{H\alpha}=3$ mag). 
On the other hand, ID 213 is not detected at 24 $\mu$m and the 3$\sigma$ upper limit of $L_\mathrm{IR}<1.7\times10^{12}~L_\odot$ is given.

\section{Discussion and summary}
\label{sec;discussion}

The separation of $\sim$30 kpc and the velocity offset of $\sim$300 km s$^{-1}$ between ID 191 and ID 193 suggest that they are probably in the initial stage of a merger.
While ID 193 is already red in the rest-frame optical and bright in the IR/radio emission, ID 191 is blue, meaning less dusty or young stellar populations.
Such red-blue pairs are also observed in many SMGs \citep{2002MNRAS.337....1I}.
Actually, the derived SFR of 886 $M_\odot$yr$^{-1}$ in ID193 is higher by a factor of about seven with respect to the $M_*$-SFR relation of normal HAEs ($main~sequence$; \citealt{2013ApJ...778..114T}) and is similar to that of SMGs, where extremely high star formation is thought to be driven by major mergers \citep{2008ApJ...680..246T,2010ApJ...724..233E,2012MNRAS.425.1320I}.
Even the high-redshift disk galaxies fed by cold accretion of gas through cosmic filaments rarely show such extremely high SFRs \citep{2009Natur.457..451D, 2008ApJ...687...59G}.
Although this classification based on colors and SFRs is not straightforward, measurements of $L_\mathrm{IR}/L'_\mathrm{CO}$ ratio are helpful for a better understanding of the star-formation mode as it reflects how molecular gas is being turned into stars.

\begin{figure}[t]
\includegraphics[width=1\linewidth]{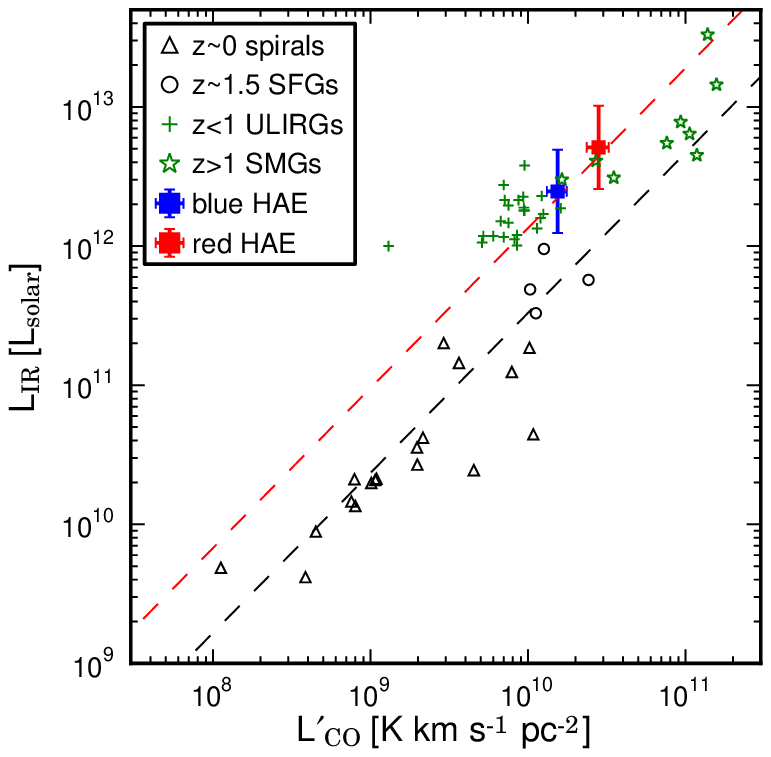}
\caption{IR vs. CO $J=1-0$ luminosities of the HAEs with CO detection along with SMGs at $z>1$ \citep[green stars:][]{2010ApJ...723.1139H,2011MNRAS.412.1913I,2013ApJ...772..137I}, ULIRGs at $z<1$ \citep[green crosses:][]{1997ApJ...478..144S}, optical/near-IR selected star-forming galaxies at $z>1$ \citep[black circles:][]{2012MNRAS.426..258A}, and local spirals \citep[black triangles:][]{2012PASJ...64...55M,2007PASJ...59..117K,2001ApJ...548..681B}.
Dashed black and red lines show the best-fitting relations for normal star-forming galaxies and luminous mergers \citep{2010MNRAS.407.2091G}. \label{fig;COIR}}
\end{figure}

In Figure \ref{fig;COIR}, we plot the two HAEs with CO and IR/radio detections on the $L_\mathrm{IR}-L'_\mathrm{CO}$ diagram to compare them with other populations taken from the literature \citep{2012PASJ...64...55M,2007PASJ...59..117K,2001ApJ...548..681B,2012MNRAS.426..258A,1997ApJ...478..144S,2010ApJ...723.1139H,2011MNRAS.412.1913I,2013ApJ...772..137I,2010MNRAS.407.2091G}.
Many previous studies at high redshift are based on the CO $J=3-2$ and $J=4-3$ line observations, leading to uncertainties in the estimates of the CO $J=1-0$ luminosities because of variations of gas excitation.
Therefore, we use only objects whose CO $J=$1--0 luminosities have been directly measured.
Both red and blue HAE have a higher IR luminosity compared to normal star-forming galaxies at a fixed CO luminosity and are situated close to the regime of low-redshift ULIRGs.
On the other hand, some SMGs seem to follow the relation of normal star-forming galaxies.
Recent analyses of galaxy morphologies in $Hubble~Space~Telescope$ images show that only 30\% of SMGs are associated with ongoing mergers \citep{2014ApJ...785..111W} while most of low-redshift ULIRGs are likely to be merger-driven starbursts \citep{2012ApJ...757...23K}.
This merger fraction of SMGs is highly contentious, with several high-resolution CO studies finding it to be 100\% \citep[e.g.][]{2010ApJ...724..233E}.
The starburst in ID 191 and ID193 is hard to explain by the first encounter or the final coalescence of a merger between them because the separation of 30 kpc is too large to induce violent star formation \citep{2009PASJ...61..481S}.
The red HAE with the high SFR, by itself, might be in a late stage of merging where the starburst is induced by a past interaction with another galaxy.
In contrast, the IR luminosity measured in the blue one is highly uncertain because of the low signal-noise ratio of 1.8$\sigma$ in the radio emission.
Therefore, we do not use ID 191 to characterize the molecular gas properties.

CO $J=1-0$ observations allow us to derive molecular gas masses independent of its excitation level.
The difference of star-formation mode is closely related to a CO to H$_2$+He conversion factor, $\alpha_\mathrm{CO}$.
Recent measurements of dust mass and dynamical mass demonstrate that $\alpha_\mathrm{CO}=3.6~M_\odot$ (K km s$^{-1}$ pc$^{-2})^{-1}$ is likely to be appropriate for normal star-forming galaxies at high-redshift \citep{2010ApJ...713..686D, 2011ApJ...740L..15M} and $\alpha_\mathrm{CO}=0.8$ is for starburst galaxies \citep{2013ARA&A..51..207B, 2013ApJ...772..137I, 1998ApJ...507..615D}.
Provided that ID193 is in the starburst mode, the molecular gas mass and the gas fraction are $M_\mathrm{gas}=2.2\times10^{10}~M_\odot$ and $f_\mathrm{gas}=M_\mathrm{gas}/(M_\mathrm{gas}+M_*)\sim31\%$, respectively.
Its gas depletion timescale is $\tau_\mathrm{depl}=M_\mathrm{gas}/\mathrm{SFR}\sim$ 25 Myr which is roughly consistent with that of SMGs at similar redshift \citep{2005MNRAS.359.1165G,2011MNRAS.412.1913I}.
Even if using $\alpha_\mathrm{CO}=3.6$, we find that $\tau_\mathrm{depl}=110~$Myr is still shorter compared to normal star-forming galaxies (400 Myr in Salpeter IMF; \citealt{2013ApJ...768...74T}).

In this work, we have discovered a close pair of CO emitting HAEs, ID 191 and ID 193, in the USS1558--003 proto-cluster at $z=2.5$.
Because the sum of their stellar masses becomes $M_*\sim10^{11}~M_\odot$ and subsequent star-formation activities would further increase it,
such a merger system can be a good candidate of a progenitor of massive quiescent galaxies in local cluster cores.
Moreover, ID 193 shows a violent star-formation activity (SFR=886 $M_\odot$yr$^{-1}$), high $L_\mathrm{IR}/L'_\mathrm{CO}$ ratio (high SFE) and red optical color.
We interpret it as a star-bursting galaxy driven by a gas-rich merger with ID 191 or another galaxy.
Given the identification of a rare event with the depletion timescale of $\tau_\mathrm{depl}=25$ Myr, merger events could frequently occur in the extremely dense environment as surrounding galaxies are expected to be dragged by a gravitational potential of a large halo.
By dissipative processes between gas-rich galaxies, the systems that have undergone merger events could be observed as the remnants with little or no rotation (slow rotators) at $z=0$ \citep{2013arXiv1311.0284N}.
This hypothesis can naturally account for the observational results that slow rotators are frequently observed in local cluster cores \citep{2011MNRAS.416.1680C}.
To confirm whether gas-rich merger events are in fact common in high-density environments at $z>2$, we need a statistical sample.
The combination of a wide-field \ha narrow-band imaging with a future wide-field observation of CO $J=1-0$ emission with ALMA Band-1 ($>$7 arcmin$^{2}$) will be powerful for such studies of star-formation modes in proto-clusters at high redshift.

\

This paper is based on data collected at JVLA, which is operated by the National Radio Astronomy Observatory. 
This research has also made use of the NASA/ IPAC Infrared Science Archive, which is operated by the Jet Propulsion Laboratory, California Institute of Technology, under contract with the National Aeronautics and Space Administration.
Data analysis were carried out on common use data analysis computer system at the Astronomy Data Center, ADC, of the National Astronomical Observatory of Japan.
We thank the anonymous referee who gave us many useful comments, which improved the paper.
K.T. and Y.K. acknowledge the support from the Japan Society for the Promotion of Science (JSPS) through JSPS research fellowships for young scientists.  
T.K. acknowledges the financial support in part by a Grant-in-Aid for the Scientific Research (Nos.\, 18684004, 21340045, and 24244015) by the Japanese Ministry of Education, Culture, Sports, Science and Technology.
Y.T. acknowledges the support from a JSPS KAKENHI grant number 25103503.


\end{document}